# Physical features of accumulation and distribution processes of small disperse coal dust precipitations and absorbed radioactive chemical elements in iodine air filter at nuclear power plant

O. P. Ledenyov, I. M. Neklyudov, P. Ya. Poltinin, L. I. Fedorova

*National Scientific Centre Kharkov Institute of Physics and Technology, Academicheskaya 1, Kharkov 61108, Ukraine.*

The physical features of absorption process of radioactive chemical elements and their isotopes in the iodine air filters of the type of *АУ-1500* at the nuclear power plants are researched. It is shown that the non-homogenous spatial distribution of absorbed radioactive chemical elements and their isotopes in the iodine air filter, probed by the gamma-activation analysis method, is well correlated with the spatial distribution of small disperse coal dust precipitations in the iodine air filter. This circumstance points out to an important role by the small disperse coal dust fractions of absorber in the absorption process of radioactive chemical elements and their isotopes in the iodine air filter. The physical origins of characteristic interaction between the radioactive chemical elements and the accumulated small disperse coal dust precipitations in an iodine air filter are considered. The analysis of influence by the researched physical processes on the technical characteristics and functionality of iodine air filters at nuclear power plants is completed.

PACS numbers: 45.70.-n, 45.70.Mg, 87.16.dp .

Keywords: air filtering, iodine air filter (*IAF*), air – dust aerosol, granular filtering medium (*GFM*), cylindrical coal adsorbent granules, small disperse coal dust fractions, coal dust precipitations, chemical absorber, heating ventilation and cooling (*HVAC*) systems, radioactive chemical elements (*Iodine*, *Barium*, *Strontium*), radioactive isotopes, nuclear power plant (*NPP*).

## Introduction.

The iodine air filters (*IAF*) have been used for the air filtering and cleaning from the radioactive *Iodine*, and some other radioactive chemical elements and their radioactive isotopes at nuclear power plants, preventing the radioactive contamination of surrounding environment. Some aspects of the *IAF* operation issues, including the increasing aerodynamic resistance problems, were extensively researched in [1, 2]. It was found that the absorber experiences a destructive action by the air stream, resulting in the cylindrical coal granules crushing into the small disperse coal dust pieces at the sub-surface layer of absorber's input in an iodine air filter. The small disperse coal dust fractions mix with the air, forming the air-dust aerosol, which can flow in the air channels between the cylindrical coal granules in the absorber. As a result, the functional performance of iodine air filter rapidly degrades over time, because of the sharp increase of its aerodynamic resistance. In [3], it was established that the distribution of small disperse coal dust precipitations inside the granular filtering medium has a certain ordered structure with a number of clearly identified density maximums in the small disperse coal dust precipitations concentration dependence. The coal dust masses concentration maximums correspond to the accumulation of small disperse coal dust particles with the different characteristic dimensions, and they are situated at the different distances from the absorber's input surface layer. These maximums are similar to the maximums, which are normally observed in the chromatographic columns at the processing of mixed chemical substances with the complex molecular composition [4]. In the last case, the separation of mixed chemical substances takes place, depending on their mass and velocity of movement of molecules through the chromatographic column. However, there are some significant distinctions. In the iodine air filter over the *15* years of operational cycle, the source of dust continuously generates the small disperse coal dust in distinction from the field of chromatography, where the source of substance provides the mixed chemical substances with the complex molecular composition for the processing in the chromatographic columns over the short time only. This research is focused on the understanding of the nature of interaction between the air – dust aerosol stream flows and the granular filtering medium formed by the cylindrical coal granules, which is important for the optimization of absorption process of radioactive chemical elements and their isotopes in the iodine air filters of the type of *АУ-1500* at the nuclear power plants. The physical role of small disperse coal dust masses concentration maximums in the absorption process of radioactive chemical elements and their isotopes in the iodine air filters of the type of *АУ-1500* at the nuclear power plants is also clarified.

# Small disperse coal dust fraction structure and its transportation properties in iodine air filter.

The cylindrical coal granules of absorber of the type *CKT-3* (length *3.2 mm*, diameter *1.8 mm*) as well as other absorbent coal materials have complex internal geometrical structure. They consist of the numerous absorption pours with the different diameters – from a few angstroms up to submicron dimensions. The characteristic distribution dependence of these pours as a function of their dimensions has a number of narrow maximums [5]. The corresponding characteristic dependence of the total volume of pores on the logarithm of pore's radius *V(log r)* in an absorber of the type of *CKT-3* in an iodine air filter is shown in Fig. 1.

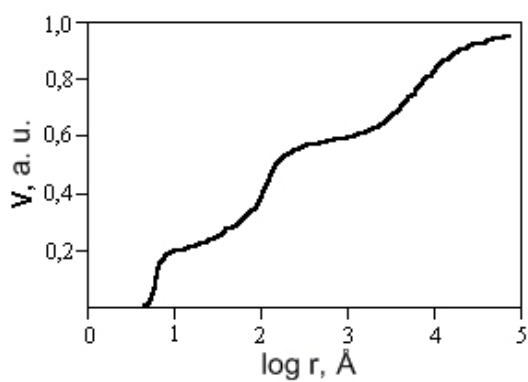

*Fig. 1. Characteristic dependence of total volume of pores on logarithm of pore's radius V(log r) in absorber of type of CKT-3 in iodine air filter.*

This graph is obtained by the method of cumulative summation of the dependence of the derivative $dV/d(\log r)$ on the logarithm of pore's radius $\log r$ [5]. The important problem on the absorption properties in the pores will be discussed later. Let us note that the given internal structure of cylindrical coal granules in the absorber, which is characterized by the presence of series of pores of selected dimensions, results in an appearance of the collection of small disperse coal dust particles with the discrete dimensions. The standard process of forced dispersion of cylindrical coal granules, conducted by the method of rolling arbor, results into the derivation of mix of small disperse coal dust fractions with the different discrete dimensions rather than with the monotonously changing dimensions. This physical feature of cylindrical coal granules destruction is characteristic for a number of dust creation methods probably. For example, in the case of cylindrical coal granules destruction in the sub-surface layer of granular filtering medium at the absorber's input during the air stream flow in an iodine air filter. The characteristic dependence of distribution of small disperse coal dust particles as a function of their dimensions, obtained at the movement of dust precipitations from the left side to the right side at the slightly tilted rough substrate under the action of vibration, is shown in Fig. 2.

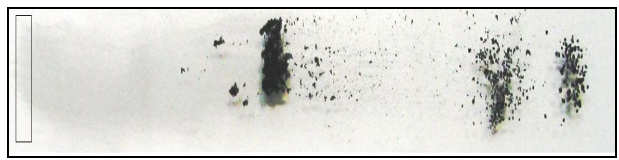

*Fig. 2. Small disperse coal dust trace on substrate at movement by dust precipitations from left side to right side at slightly tilted substrate at action of vibration (initial position is shown as rectangle at left side of substrate).*

As it can be seen in Fig. 2, during the air blow process through the iodine air filter, the group of small disperse coal dust particles is divided on the following fractions, depending on the particles transportation velocities: 1) the fraction with the large size particles with the particles diameter *d ~ 100 μm* (the coal dust precipitations on the right side of substrate); 2) the fraction with the medium size particles with the particles diameter *d ~ 30 μm* (the coal dust precipitations in the middle of substrate); 3) the fraction with the small size particles with the particles diameters *d ~ 10 μm* and *d ~ 1 μm* (the coal dust precipitations on the left side of substrate); and 4) the fraction with the very small size particles with the particles diameters from *d ~ 1 μm ... 1 nm* (the coal dust precipitations on the left side of in the beginning of substrate) It is necessary to emphasis that there are some very small disperse coal dust fractions in the dust masses, but they are not visible optically.

Thus, during the air filtering process, the group of small disperse coal dust particles in the air – dust aerosol is separated on the fractions with the various coal dust particles dimensions in the granular filtering medium in the absorber. This physical process originates, because of the fact that, in the case, when the mean free path of molecules of air *l* is smaller than the radius of spherical particle *r*, then the magnitude of capturing force, acting on the particle, is defined by the magnitude of *Stocks* force

$$F_S = 6\pi\mu r \upsilon, \qquad (1)$$

where $\mu$ is the air viscosity, $\upsilon$ is the air velocity in relation to the dust particle velocity. The expression for the acceleration, experienced by the small disperse coal dust particles, can be written as $a = F_S/m$, where *m* is the mass of a coal dust particle. The dependence of the coal dust particle's acceleration on the coal dust particle's radius is presented in eq. (2)

$$a = \frac{9\mu\upsilon}{2\rho r^2}, \qquad (2)$$

where $\rho$ is the density of particle's material. Taking to the consideration the fact that the mean velocity of particle is proportional to the acceleration of particle, hence the transportation velocity of coal dust particle is a function of the particle's dimension. Therefore, the smaller particles have bigger transportation velocities in the absorber in an iodine air filter.

In [3], it was shown that the small disperse coal dust is separated on the fractions, depending on the velocities of dust particles movement through the granular filtering medium in an absorber. The distribution of maximums of dust masses densities is similar to the well known distribution of molecules in the field of chromatography. In the researched case, the maximums of small disperse coal dust masses density are observed in the conditions, when the small disperse coal dust has been generated continuously; whereas, in the field of chromatography, the general assumption is that the accurate maximums can be obtained in the case of use of a single non-renewable source of complex molecular substance with the smaller thickness than the length of chromatographic column. Let us emphasis that, in the researched case, it was found that there is a continuous-time processes of small disperse coal dust generation, accumulation and maximums creation in the absorber of an iodine air filter. The generated small disperse coal dust has a number of fractions of coal dust particles with the certain dimensions. Nevertheless, this circumstance has no any particular effect on the shape of curves, approximating the maximums, and it doesn't lead to the full disappearance of maximums. This fact can be well understood, taking to the consideration that the velocity of small disperse coal dust transportation depends on the nature of coal dust particles scattering on the other coal dust particles rather than on the character of coal dust particles interaction with the cylindrical coal granules. Indeed, in the first approximation, a collision between the coal dust particles of different dimensions and the cylindrical coal granule can be described as an elastic collision between the particles with the very different masses. During the particles collision process, the transverse (to the collision boundary) impulse changes to the opposite sign impulse; but the longitudinal (to the collision boundary) impulse preserves the constant magnitude. In this case, there is the constant mass stream flow along the air channel, where the particle is transported, and there is no the "de-acceleration" of air –dust aerosol stream flow. The dissipation of impulse and energy of coal dust particles has place at the scattering of coal dust particles on the cylindrical coal granules, because of both the non-ideal geometrical forms of coal dust particles and the possibility of their deformation and destruction. However, the collisions between the coal dust particles are far more important.

In this case, the mean free path by coal dust particles differs from the length of air channel between the cylindrical coal granules $l_C$, and the mean free path is equal to the distance between the coal dust particles in air – dust aerosol $l_P$. This distance depends on the density of coal dust particles $n$, and it can be represented as $l_P \approx V/n^{1/3}$, where $V$ is the volume taken by the air – dust aerosol. The characteristic magnitude of diffusion coefficient at the velocity of coal dust particles $\upsilon_P$ will be equal to $D_P = l_P \upsilon_P \sim 1/n^{1/3}$. Let us note that, in the considered simple case, the coefficient of diffusion will decrease at an increase of density of coal dust particles of given type. In our approach, the coal dust particles of various dimensions and masses don't create the effective scattering centers for each other, hence the presence and the high density of particles of same dimension is important [6]. The collisions with the coal dust particles of significantly bigger dimension and masses as well as with the cylindrical coal granules of an absorber don't result in the impulse transfer and have no effective influence on the small disperse coal dust particles transportation in an iodine air filter. As it was shown in [2], the small disperse coal dust particles of all the dimensions, created in the sub-surface layer at the absorber's input, are captured and transported by the air stream during the air blow process in a working iodine air filter. As it follows from the eq. (2), the small disperse coal dust particles of small dimensions get the high velocities of transportation; and the small disperse coal dust particles of bigger sizes have small velocity of transportation during the air – dust aerosol flow through an absorber. However, it is necessary to mention that, the small size dust particles with the high velocities, appearing at the subsequent stages of iodine air filter operation, can reach the position of big size dust particles with low velocities, penetrate through the big size dust particles clot and experience the small decrease of their own velocity. Whereas, the coal dust particles with the same dimensions experience the sharp decrease of their velocity, and form the dust clot in which all the coal dust particles have the similar dimensions. At an increase of packaging density of small disperse coal dust particles in the dust clot, the diffusion coefficient and averaged velocity of transportation of these particles decreases. Thus, the origination of creation of dust clots, consisting of small disperse coal dust particles with the different dimensions, takes place. During this process, the big dust particles fractions have less mobility comparing to the small dust particles fractions, hence during the air blow process, the big particles fractions form the dust layer consisting of coal dust particles with the all the possible dimensions. There are the particles with the very small dimensions as well as the particles with the big dimensions in this layer with the accumulated small disperse coal dust. The small disperse coal dust structurization process can be originated in the granular filtering medium, if the certain density of coal dust masses is reached [7].

### Correlations in distribution of small disperse coal dust precipitations and absorbed chemical elements in iodine air filter.

As it was discovered in [3], the distribution of coal dust masses, obtained in the granular filtering medium of an absorber of the type of *CKT-3* at the modeling experiment, has the dependence, shown in Fig. 3.

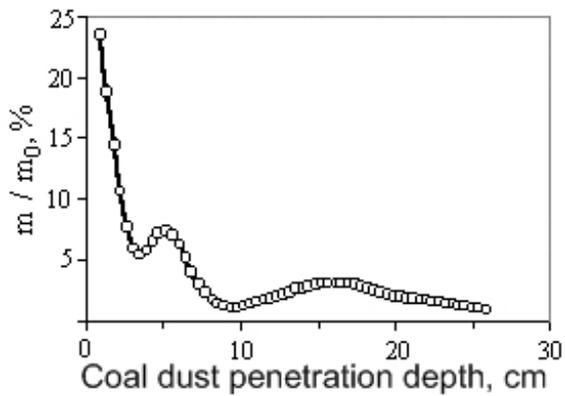

*Fig. 3. Distribution of relative mass of small disperse coal dust fractions as function of distance from absorber's input surface in modeling experiment (after [3]).*

Let us note that the distribution of small disperse coal dust particles across the thickness of coal dust layer in an absorber of the type of *CKT-3* in an iodine air filter at the nuclear power plant over many years of *IAF's* operation was not researched so far. However, the detailed researches on the distribution of *Iodine*, *Barium* and *Strontium*, including some other chemical elements and their isotopes, were conducted with the application of the gamma - activation analysis method [8]. With this purpose, the samples of granular filtering medium (the kernels) were taken by the special probes in the two specified positions along the length of absorbent layer: 1) in the center of absorber, and 2) at the periphery of absorber.

As it was found, the distribution of chemical elements and their isotopes had the dependences with the clearly defined maximums, which were not explained in [8]. The curves, obtained in [8], are shown in Figs. 4, 6 and 7.

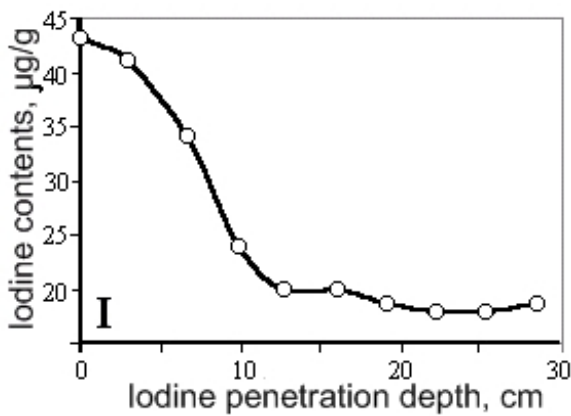

*Fig. 4. Distribution of mass content of Iodine (I) as function of distance from absorber's input surface in accordance with gamma - activation analysis method (after [8]).*

As it can be seen in Fig. 4, the *Iodine* is non-monotonously distributed along the absorber's length in an iodine air filter. There is a clearly defined maximum in distribution of mass density of *Iodine* (*I*) as function of distance from absorber's input surface in the beginning of absorber's length, which decreases to the minimum on the approximate distance of ~ *10 cm* from the absorber's input surface in an iodine air filter. The weak maximums can also be seen on the distance of *15 cm* from the absorber's input surface and at the absorber's output surface. The positions of *Iodine* concentration maximums will well correlate with the positions of small disperse coal dust precipitations concentration maximums in the case of introduction of increased exposition time (the exposition time has to be increased on the order of magnitude approximately), going from the results in Fig. 3 [3]. Then, the second maximum, centered at the distance of *7 cm*, will be shifted inside the absorber in an iodine air filter. It is possible to place the second maximum at the approximate distance of ~ *15 cm* by making the necessary selection of time. The third maximum, which is situated on the distance of *15 cm* will almost be shifted outside the absorber so that its initial part will still be placed in the absorber. This discussed distribution is presented in Fig. 5.

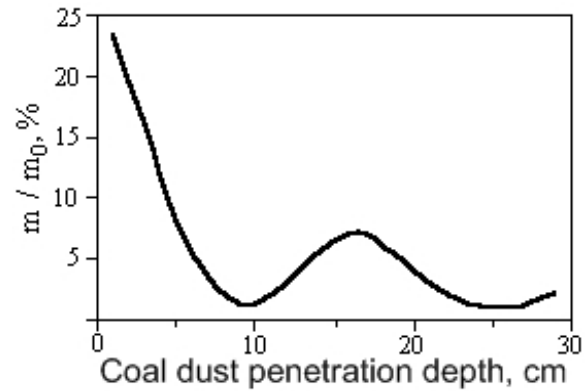

*Fig. 5. Computed distribution of relative mass of small disperse coal dust fractions as function of distance from absorber's input surface, obtained with the use of research results in [3].*

In Fig. 6, the experimental data on the distribution of *Barium* (*Ba*) as a function of distance from absorber's input surface in accordance with gamma - activation analysis method are provided [8]. As it can be seen, the *Barium* is distributed strongly non-homogeneously, but the positions of radioactive *Barium* concentration maximums correlate with the positions of small disperse coal dust precipitations concentration maximums. Let us emphasis that the radioactive *Barium* and *Iodine* are distributed strongly non-homogeneously, but their maximums correlate with the positions of small disperse coal dust precipitations concentration maximums in Fig. 5.

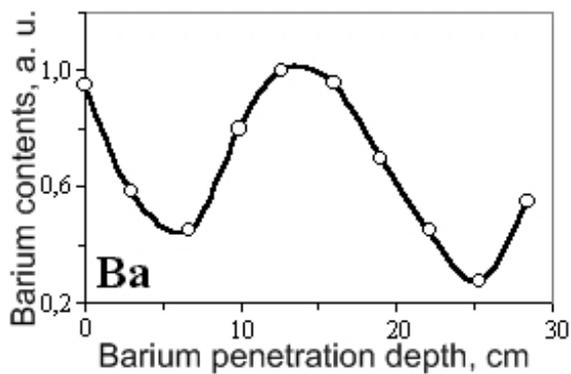

*Fig. 6. Distribution of Barium (Ba) as function of distance from absorber's input surface in accordance with gamma - activation analysis method (after [8]).*

In Fig. 7, the experimental data on the distribution of *Strontium* (*Sr*) as a function of distance from absorber's input surface in accordance with the gamma - activation analysis method are presented. As it can be seen, the *Sr* distribution of is non-monotonous, and the positions of concentration maximums correlate with the positions of small disperse coal dust precipitations concentration maximums similar to the cases of the *Iodine* and *Barium*. However, there are some distinctions between these distributions, which can be explained by the features of interaction of the chemical elements with the cylindrical coal granules and coal dust fractions, which will be discussed below. Let us note that the distributions of all these radioactive chemical elements were researched by the gamma - activation analysis method with the use of the kernel, taken at the periphery of granular filtering medium in an iodine air filter.

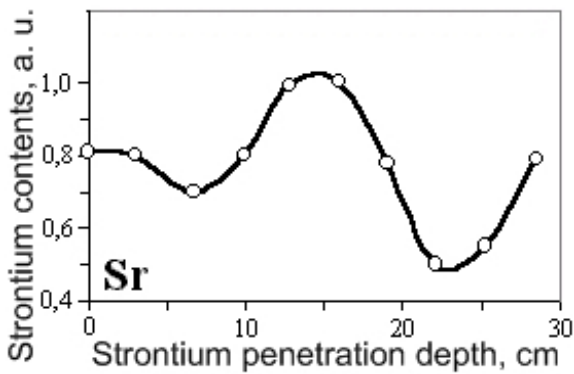

*Fig. 7. Distribution of Strontium (Sr) as function of distance from absorber's input surface in accordance with gamma - activation analysis method (after [8]).*

In the case of the kernel, taken at the center of granular filtering medium in an iodine air filter, the slightly different experimental data on the distribution of *Strontium* as a function of distance from absorber's input surface in accordance with the gamma - activation analysis method were obtained. In the researched absorber with the diameter 1 *m* , there is the slightly different air – dust aerosol stream flow through the various parts of absorber due to the physical features of small disperse coal dust generation in the various parts of granular filtering medium. Therefore, the masses concentration maximums in the distributions of the chemical elements as a function of the distance from the absorber's input surface in accordance with the gamma - activation analysis method are positioned at the different distances at the absorbent layer thickness length.

## Discussion on research results.

The comparative analysis of data, including: 1) the results of modeling experiments toward the research on the distribution of the relative mass of small disperse coal dust fractions as a function of the distance from the absorber's input surface, and 2) the results of the distribution of *Iodine* , *Barium* , *Strontium* as the functions of distances from the absorber's input surface in accordance with the gamma - activation analysis method, show that the distributions of both the small disperse coal dust particles and the chemical elements are well correlated. The contents of the *Iodine*, measured in the cylindrical coal granules at the distance ($L \approx 25$ cm), where there is almost no the small disperse coal dust fraction, approaches to *18 μg/g*. Assuming that the variation of the contents is fully connected with the coal dust mass, we will get the approximate contents of *Iodine* in the initial layer of granular filtering medium in an iodine air filter, with is ~ *100μg/g*. Thus, the small disperse coal dust has the much stronger *Iodine* absorption properties in comparison with the cylindrical coal granules absorption properties. This is connected with the fact that the active absorbent surface is increased during the cylindrical coal granules destruction process, hence the chemical elements and their isotopes can penetrate inside the absorbent material, getting access to the small pores, where there is most intensive absorption process. In addition, the activation energy, connected with the absorption process, has an increased magnitude in the small pores of absorbent material [9]. In agreement with the dependence in Fig. 4, almost all the *Iodine*, connected with the small disperse coal dust, is absorbed at the first coal masses concentration maximum, hence the contents of the *Iodine* in the second coal dust mass concentration maximum at the distance $L \approx 15$ cm is almost the same as the contents of the *Iodine* in the cylindrical coal granules. The situation is different in the cases of the *Barium* and *Strontium*. As it can be seen in Figs. 6 and 7, the first coal dust masses concentration maximum can not be regarded as a main absorber of these chemical elements, whereas the second coal dust masses concentration maximum absorbs the considerable amounts of the *Barium* and *Strontium*. Let us take to the consideration the fact that the small quantity (*7%*) of relative excessive mass of small disperse coal dust is concentrated in the second dust masses concentration maximum only, but the big quantity (*70%*) of the *Barium* and the big quantity (*50%*) of the *Strontium* are absorbed in the second dust mass concentration maximum in an iodine air filter. The simple comparison shows that the effectiveness of absorption of the second coal dust masses concentration maximum is approximately in the *14-15* times bigger in the case of the *Barium* and in the 30 times bigger in the case of the *Strontium* in comparison with the absorption by the cylindrical coal granules. The relatively small magnitude of absorption of the *Barium* and *Strontium* at the first coal dust masses concentration maximum is probably connected with the fact that the *Barium* and *Strontium* are substituted by the *Iodine*. It can be well noticed that the *Barium* and *Strontium* are effectively absorbed at the third coal dust masses concentration maximum, which almost shifted outside the absorber in an iodine air filter.

### Conclusion.

The research on the physical features of absorption process of chemical elements and their isotopes in the iodine air filters of the type of *АУ-1500* at the nuclear power plants is completed. It is shown that the non-homogenous spatial distribution of absorbed chemical elements and their isotopes in the iodine air filter, probed by the gamma-activation analysis method, is well correlated with the spatial distribution of small disperse coal dust precipitations in the iodine air filter. For instance, the obtained research data show that the absorption of the *Iodine* and other chemical elements mainly takes place in the small disperse coal dust precipitations, which are created and distributed in the form of coal dust masses concentration maximums in the granular filtering medium in the iodine air filter. These coal dust masses concentration maximums can shift, starting at the absorber's input surface and going deeper into the granular filtering medium in the iodine air filter. It is necessary to comment that the *Iodine* is absorbed by the first coal dust masses concentration maximum, which is not shifted at an action by the air – dust aerosol stream flow, hence the *Iodine* has small mobility and tends to be accumulated in the iodine air filter of the type of *АУ-1500* at the nuclear power plants. There is a high probability that the accumulated *Barium* and *Strontium* can be jettisoned into the atmosphere at a certain stage of operation of the iodine air filter of the type of *АУ-1500* at the nuclear power plants, because of the observed characteristic properties of small disperse coal dust spatial distribution, accumulation, and structurization. This circumstance points out to an important role by the small disperse coal dust fractions in the absorption process of chemical elements and their isotopes in the iodine air filter. The physical origins of characteristic interaction between the chemical elements and the accumulated small disperse coal dust precipitations in the iodine air filter are researched. The analysis of influence by the researched physical processes on the technical characteristics and functionality of iodine air filters at the nuclear power plants is completed. The obtained research results will be used to make an optimization of technical characteristics and create the advanced designs of the iodine air filter of the type of *АУ-1500* at the nuclear power plants.

Authors thank P. A. Khaimovich for the small disperse coal dust objects photography.

This innovative research is completed in the frames of the nuclear science and technology fundamental research program at the National Scientific Centre Kharkov Institute of Physics and Technology (*NSC KIPT*).

This research paper was published in the *Problems of Atomic Science and Technology* (*VANT*) [10].

*E-mail:  ledenyov@kipt.kharkov.ua

———————